\documentclass[a4paper]{jpconf}
\usepackage{graphicx}
\begin{document}
\title{Dynamics of the topological structures in inhomogeneous
  media\footnote{Talk given by WJZ at the 5th International Symposium
on Quantum Theory and Symmetries,Valladolid, Spain. To be published in
  the Proceedings in a special issue of Journal of Physics A.}}

\author{LA Ferreira$^1$, Bernard Piette$^2$ and WJ Zakrzewski$^2$ }

\address{$^1$Instituto de F\'isica de S\~ao Carlos, IFSC/USP, S\~ao Carlos, SP, Brazil}
\address{$^2$Department of Mathematical Sciences,  University
of Durham, Durham DH1 3LE, U.K}

\ead{laf@ifsc.usp.br, B.M.A.G.Piette@durham.ac.uk, W.J.Zakrzewski@durham.ac.uk}

\begin{abstract}
We present a general review of the dynamics of topological solitons in 1 and 2 dimensions
and then discuss some recent work on the scattering of various solitonic objects (such as kinks and breathers etc) 
 on potential obstructions.

\end{abstract}
\section{Introduction}

Topological solitons arise in many areas of applied mathematics
and in the mathematical description of some processes in physics - for a good review see \cite{Paul}.

They arise in the mathematical description of objects that are localised in space 
as the energy density of these objects is nonzero only in a finite region; {\it ie}
it is significantly nonzero in a small region and goes to zero, exponentially
or as an inverse power, as one moves away from this region. 

Their stability is guaranteed by topological considerations, normally
associated with the topology of $S^N\rightarrow S^N$ maps.

The simplest examples of such maps (in 1+1 and 2+1) dimensions involve, respectively,
 the Sine-Gordon kinks and the solitons based on sigma models. 

 As is well known the Lagrangian density of the (1+1) dimensional sine-Gordon model is given by
\begin{equation}
\label{lagrangianSG} 
{\cal L}=\partial_{\mu}\phi\cdot\partial^{\mu}\phi
\,-\,\lambda \sin^2(\phi).
\end{equation}

Solutions of the equations of motion which follow from  (\ref{lagrangianSG})
are well known. They involve kinks and antikinks, which are topological
solitons; breathers, which can be thought of as bound states
of kinks and antikinks, further bound states of kinks and breathers, 
and many other solutions, less interesting from our point of view.

The Lagrangian for the (2+1) dimensional sigma model is given by:
\begin{equation}
\label{lagrangian} 
{\cal L}=\partial_{\mu}\vec{\phi}\cdot\partial^{\mu}\vec{\phi}
-\theta_{S}\left[(\partial_{\mu}\vec{\phi} \cdot
\partial^{\mu}\vec{\phi})^{2} -(\partial_{\mu}\vec{\phi} \cdot
\partial_{\nu}\vec{\phi}) (\partial^{\mu}\vec{\phi} \cdot
\partial^{\nu}\vec{\phi})\right] -  V(\vec{\phi}),
\end{equation}
where for $V$ we can take any `simple' function of $\phi_3$ which vanishes as
$\phi_3=1$. So, in this work we take
\begin{equation}
V(\vec{\phi}) = \mu(1-\phi_3^2).
\label{pot}
\end{equation}
Furthermore, we require that the vector $\vec{\phi}$ lies on the unit sphere ${\cal S}^{2}$
hence $\vec{\phi}\cdot\vec{\phi}=1$. These last two conditions make the static finite energy
solutions of the Euler Lagrange equations which follow from (\ref{lagrangian}), the solitons, topological
and based on the topology of $S^2\rightarrow S^2$ maps.
The requirement that the static energy is finite forces the field $\phi_3$ to go to +1 at spatial
infinity thus compactifying $R^2$ and allowing us to consider $\vec{\phi}$ 
on the extended plane $R^2\bigcup{\infty}$  topologically equivalent to  ${\cal S}^{2}$.

The three terms in (\ref{lagrangian})
are, from left to right, the pure ${\cal S}\sp2$ sigma model, the Skyrme and the potential term.
The last two terms are needed to stabilize the solitons. As long as they are small they have no influence
on the topology of the fields.

Incidentally, other potential terms (\ref{pot}) have also been studied \cite{weidig}.
The results, in these cases, are only slightly different but their generic features are the same. 
Hence in this work we discuss our results obtained for (\ref{pot}).
In 3 spatial dimensions we have skyrmions and monopoles but 
they will not be discussed here; the interested reader can find many interesting details in \cite{Paul}.

\section{Dynamics}

The dynamics of the Sine Gordon kinks is well known:
the kinks reflect from each other and, in fact, not much can happen as the motion
is in one dimension.

The dynamics of the two dimensional sigma model solitons can be either, 
 relativistic, ({\it i.e.} based on the Lagrangian above)
or based on the Landau-Lifshitz equation which is given by 
\begin{equation}
{\partial \vec{\phi}\over \partial t}\,=\, \vec{\phi}\,\times \, {\partial L\over \partial \vec{\phi}}
\end{equation}
where  $L$  stands for the spatial part of ${\cal L}$, {\it i.e.} of (\ref{lagrangian}).

The dynamics of both cases is very different.
In the relativisitic case we have the familiar $90^0$ scattering. Thus, when two solitons are
 sent towards each  other head on ({\it i.e.} at a zero impact parameter), the system evolves 
in such a way that after the scattering we have two outgoing solitons which are moving in the direction
perpendicular to the motion of the original solitons.

This has been explained in many ways; the most compelling one involves the indistinguishability
of solitons \cite{Paul}. As the system of two solitons is described by a function which is symmetric
with respect of the interchange of the positions of these solitons,  the phase space
of their relative position, is really described by $R^2$ mod a reflection
with respect to the line joining their positions. Hence, effectively, the space is 
${R^2\over Z_2}$, where $Z_2$ describes this reflection, and so it is a cone.
A straight line motion in this space through the vertex of the cone corresponds to
 a $90^0$ motion when viewed in the `opened-up' $R^2$. 

In the nonrelativistic case the situation is different as the equations involve
 only first order time derivatives. Hence the motion 
takes place in a lower dimensional phase space.

This has been analysed in some detail by Papanicolaou and Tamaras \cite{nikos}
who showed that when one introduces, for a system of two solitons, 
 $\vec{r}=(x_1,x_2)$ - a 2 dimensional vector describing their relative position,
 the corresponding relative momentum $\vec{p}$ satisfies
\begin{equation}
 p_i\,\sim \alpha\,\epsilon_{ij} x_j.
\end{equation}
Hence the equation motion is of the form
\begin{equation}
{d^2x_i\over dt^2}\,\sim \alpha \epsilon_{ij}{dx_j\over dt}
\end{equation}
resulting in a motion along a circle.

 The dynamics of solitons in 3 spatial dimensions is even more complicated \cite{Paul}; however,
many aspects of it can be related to the dynamics in 2 dimensions. 

All this discussion concerned solitons moving in a free space, {\it i.e.} in a space with no potential 
obstructions. In the next section we summarise our recent results on the scattering of individual solitons 
on a spatial obstruction - either in the shape of a potential bump or a hole.

\section{Potential Obstruction}

To introduce a potential obstruction we note that the energy of the soliton field, strictly 
speaking, is
never zero, even though it vanishes exponentially as we move away from
soliton's position. Hence the obstruction has to be introduced in such a way that
it does not change the ``tail'' of the soliton {\it i.e.} it has to
vanish when, in one dimension, $\phi=0$ or $\pi$  or, in two dimensions,
when $\phi_3=1$.

\subsection{Sigma Models}

Although a choice of obstruction is highly nonunique, in our studies,
 we added an extra term 
to the Lagrangian and we chose it of the form  $\alpha(1-\phi_3^2)$ ({\it i.e.}
like (\ref{pot}) with $\alpha$ nonzero in some region of $x$
and $y$). We chose this term to be independent of $y$ so that it represented an 
obstruction,  located in some finite
region of $x$ resembling a trough in the ``hole''
case or a dam in the ``barrier'' case. Then sending the soliton from
a point far away from this obstruction we
can study its scattering on this obstruction.

We have performed many numerical simulations of such systems, varying
both the sign and value of $\alpha$ and the velocity of the incoming solitons.
In the barrier case we have found that the scattering is very elastic with
the soliton behaving like a point particle, depending on its kinetic
energy - either going over the barrier or being reflected by it.
Hence the velocity of the outgoing soliton is always very close, in magnitude,
to the original velocity.

For the hole case the situation was different.
Depending on the value of the velocity solitons were either tranmitted
or trapped and sometimes were even reflected.   In fig 1. we present plots of the
solitons' positions (in $x$ as $y=0$)
as a function of time seen in some simulations ({\it i.e.} started with 
different values of initial velocities).


\begin{figure}[h]
\begin{minipage}{12pc}
\includegraphics[width=12pc]{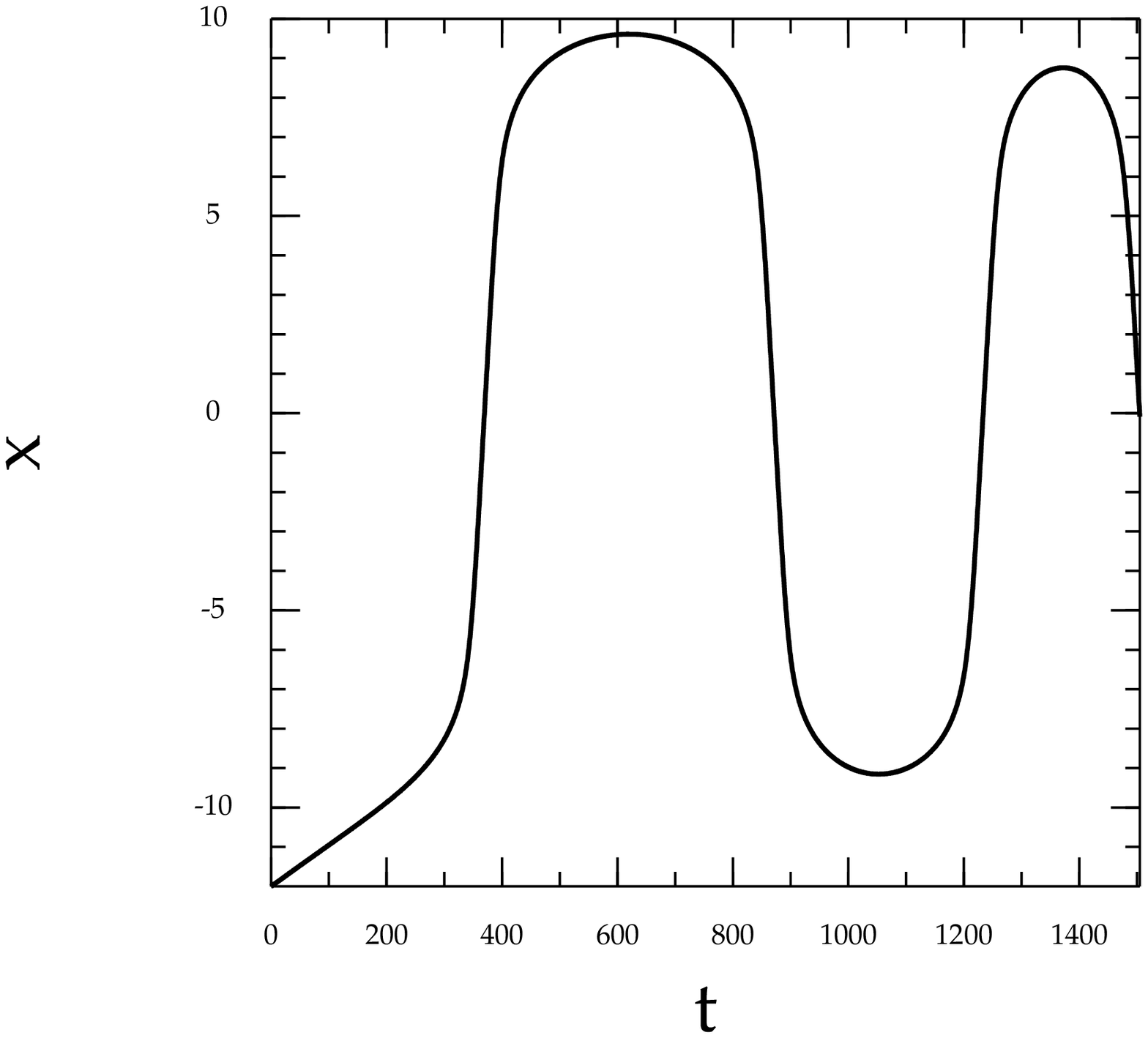}
\end{minipage}\hspace{1pc}%
\begin{minipage}{12pc}
\includegraphics[width=12pc]{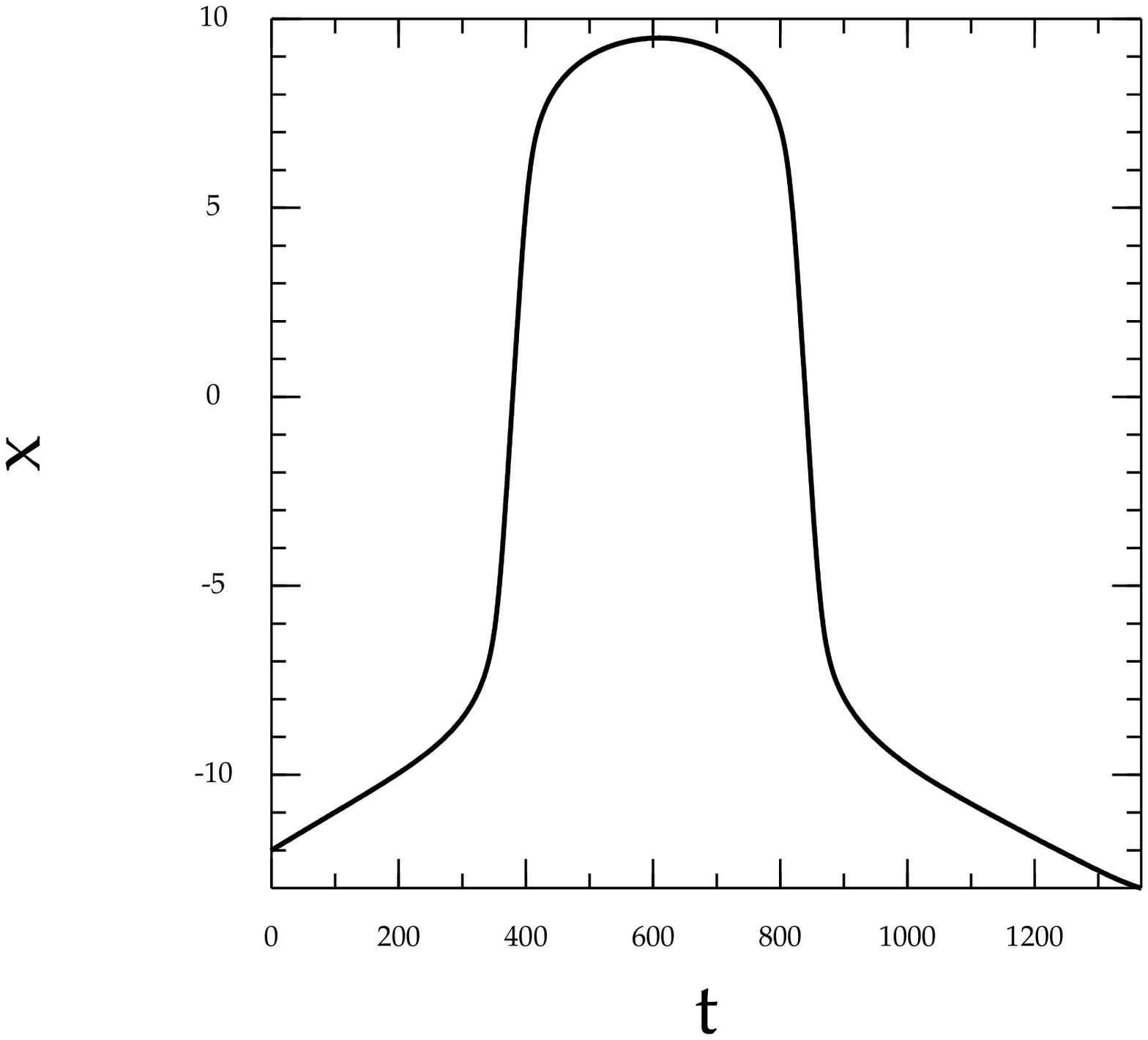}
\end{minipage}\hspace{1pc}
\begin{minipage}{12pc}
\includegraphics[width=12pc]{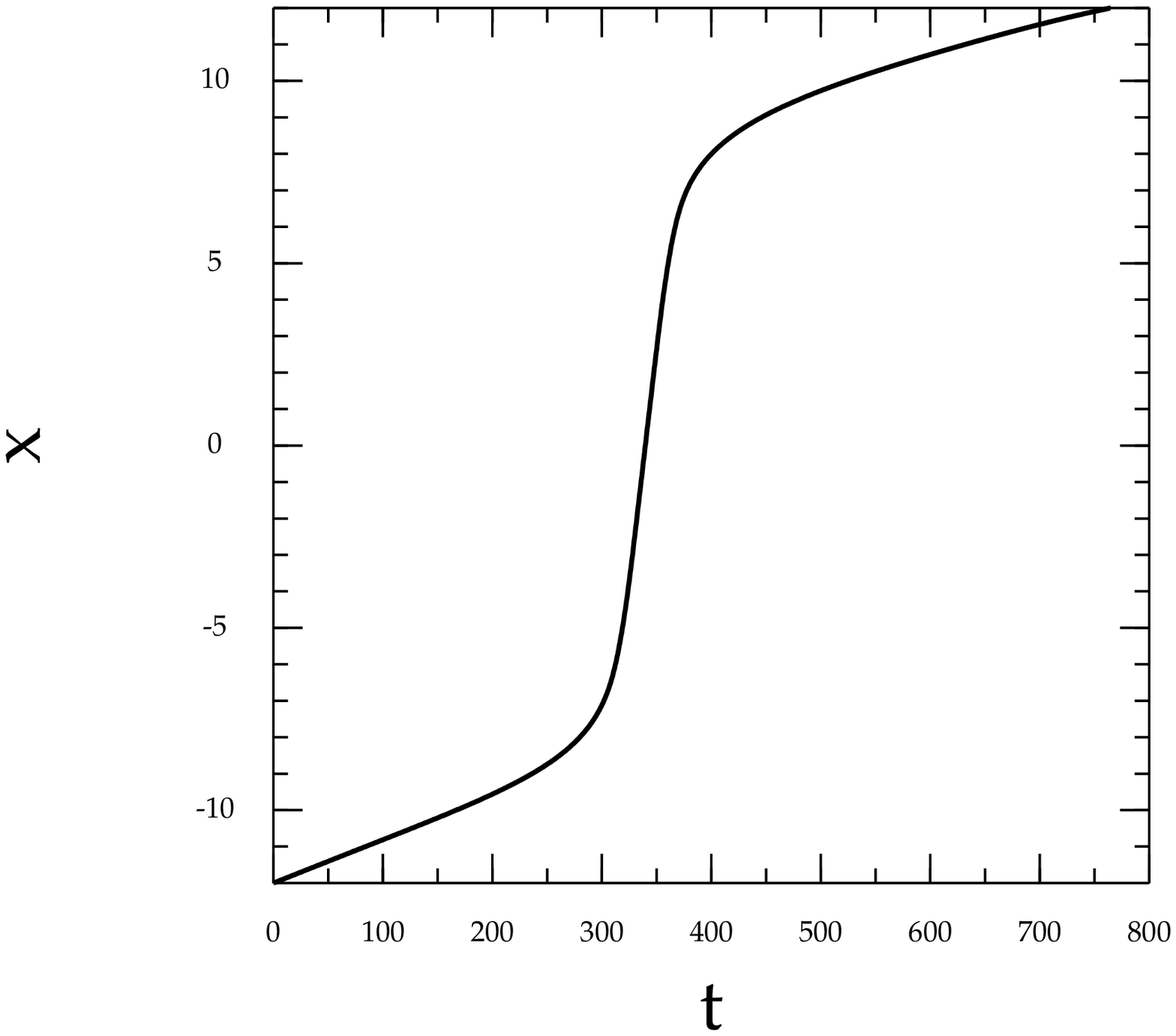}
\end{minipage}
\caption{Trajectory of a soliton during the scattering for 
a well of width $L=10$ and depth $0.2$, ($\mu$ reduced from 0.25 to 0.2). Here $\theta_s=0.25$ and a) $v=0.0106$, b) $v=0.0102$, c) $v=0.012$, respectively}
\end{figure}

In addition to the trapping, which shows that, this time, the soliton does not behave like a point
particle, we have also seen reflections!
We have studied this process in great detail and in  
\cite{sigma} we argued that the interaction of the soliton with the 
hole proceeds through the excitement
of the vibrational modes of the soliton.  This has led us to look at other models 
and, in particular, at the Sine Gordon model in 
one dimension, which does not have any genuine vibrational modes.  We discuss the results in this
model in the next section.

\subsection{Sine Gordon Model}

There are two easy ways of introducing the obstruction into the Sine-Gordon model: either
by making  $\lambda$ in (\ref{lagrangianSG}) position dependent \cite{sine}
or by altering the basic metric \cite{kourosh}. Here we discuss the results reported in
\cite{sine}; the results of \cite{kourosh} are qualitatively very similar. We restrict our 
discussion to the case of the potential hole.

The results of our studies have shown that in the sine Gordon model, as in 
the sigma model in two dimensions, the solitons can get trapped, be transmitted
and bounce back. The process is inelastic and depends on the initial condition and 
 on the size and the depth of the hole.
If the initial condition of the soliton corresponds to an 
exact Sine-Gordon kink moving with a given velocity
then there is a well defined critical velocity above which the kink get transmitted
(with a certain loss of energy). Below this critical velocity the kink can be trapped
or reflected. The ranges of velocities, at which the kink is reflected
are very narrow. The value of the critical velocity decreases with the decrease of the depth
of the well and as the hole becomes narrower the number of velocity windows corresponding
 to reflections gets larger. These windows tend to be very narrow indeed.
In fig.2 we present a typical plot of the outgoing velocity as a function of the incoming velocity 
for the case of a relatively narrow well (in this case $\lambda$  in the well is lowered from 1 down to
$0.8$) and the hole is relatively narrow - {\it i.e.} a soliton fits in it about 3 times).

\begin{figure}
\begin{center}
\includegraphics[width=14pc]{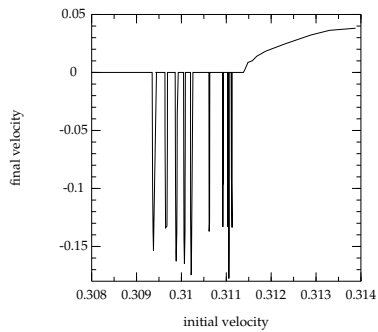}
\end{center}
\caption{\label{label1}Outgoing velocity of the kink as a function of its initial velocity}
\end{figure}
Thus we note that in the Sine-Gordon model the solitons, which 
have no vibrational modes, can still be reflected by a potential hole.
 Of course, although
the solitons have no vibrational modes they have pseudovibrational ones which can get excited and
radiate (similar conclusions,  in a different
context, were reached by Romanczukiewicz \cite{Rom}).
An example of such a pseudovibrational mode is the mode which describes 
the variation of the slope of the kink. The usual kink
solution is given by
\begin{equation}
\label{kink}
\phi(x)\,=\,2 \arctan(\exp(\theta (x-x_0)),
\end{equation}
where $x_0$ is the kink position and $\theta$ is its slope. For (\ref{kink})
to be a solution of the equation of motion, which follows from (\ref{lagrangianSG}),
we need to put $\theta$ and $\lambda$ equal. However, if we put $\theta$ different from $\lambda$
we excite the mode which corresponds to the variation of $\theta$. In fact, looking at the 
simulations in detail we have noticed that when 
the kink enters the hole for which $\lambda$ is different from $\theta$, it automatically tries 
to adjust its slope and so it excites this mode. Of course, as soon as this mode is excited
 the kink begin to radiate and it is the interaction of this radiation
with the kink itself which is reponsible for the final outcome of the scattering process.

\section{Breathers}

The Sine-Gordon model, in addition to the kink, also possesses many other solutions.
Amongst them are breathers which are  given by
\begin{equation}
\phi(x,t)\,=\,2\,\arctan\left({\sin(\omega t)\over \omega\,\cosh(\sqrt{1-\omega^2}\,x)}\right).
\end{equation}

As their energy is $16 \sqrt{1-\omega^2}$ they are often thought of as bound states
of a kink and an antikink with the binding increasing as $\omega\rightarrow 1$.
Hence it is interesting to see what happens when a breather is sent towards a hole.

We have performed several numerical experiments of such systems and we are now preparing a long
paper with  a description of our results \cite{new}. So here we will mention only some of them.
As expected our results have shown that the breathers do get trapped, pass
 with, in general, a different $\omega$ (note that increasing $\omega$ releases 
some energy), or split with either a kink or an antikink being ejected from the hole.

In fig 3 we present a couple of pictures showing a breather just before a scattering
on a hole, and some time later, in the case when the interaction with 
the hole leads to the splitting of the breather.

\begin{figure}[h]
\begin{minipage}{14pc}
\includegraphics[width=14pc]{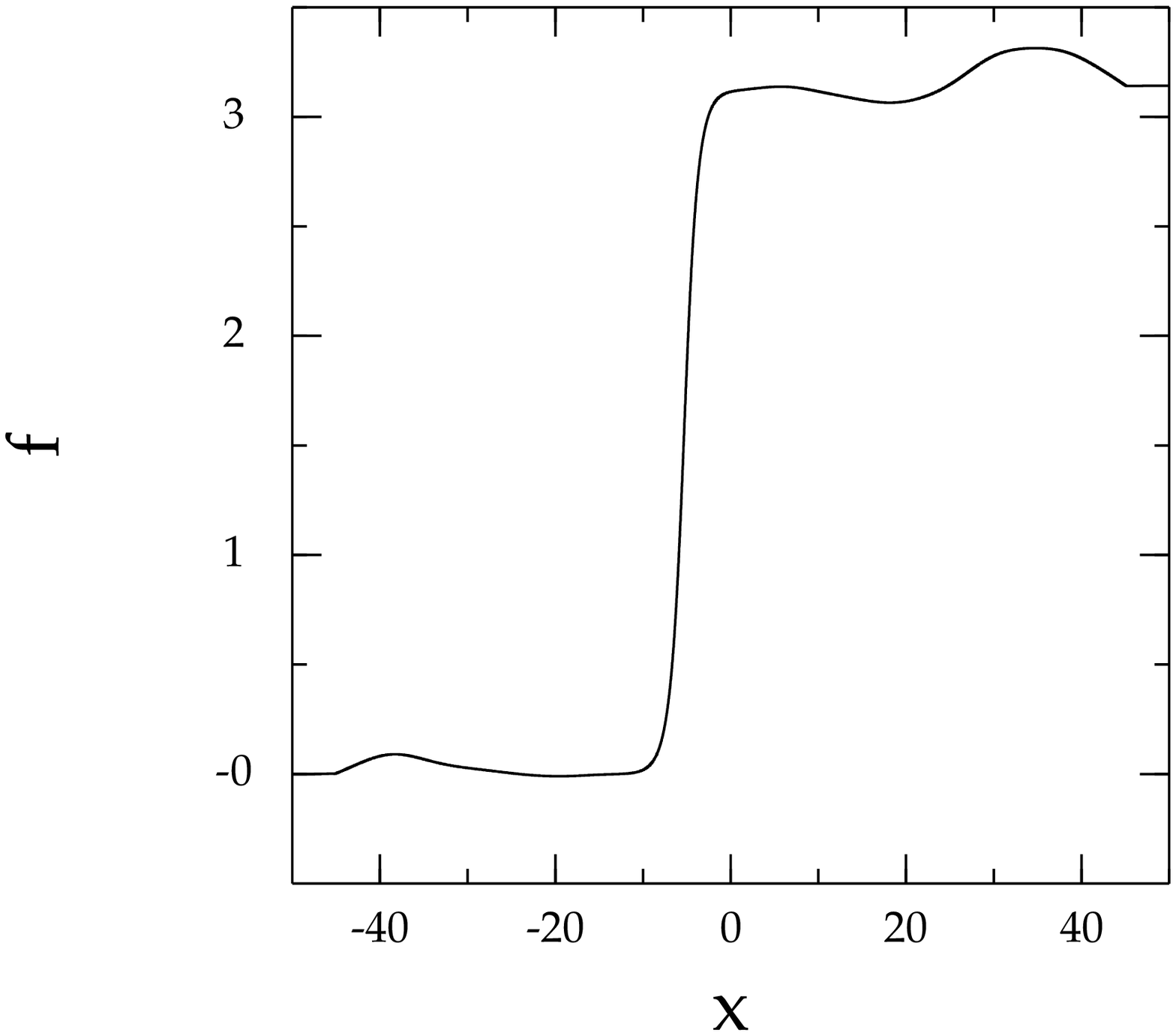}
\end{minipage}\hspace{2pc}%
\begin{minipage}{14pc}
\includegraphics[width=14pc]{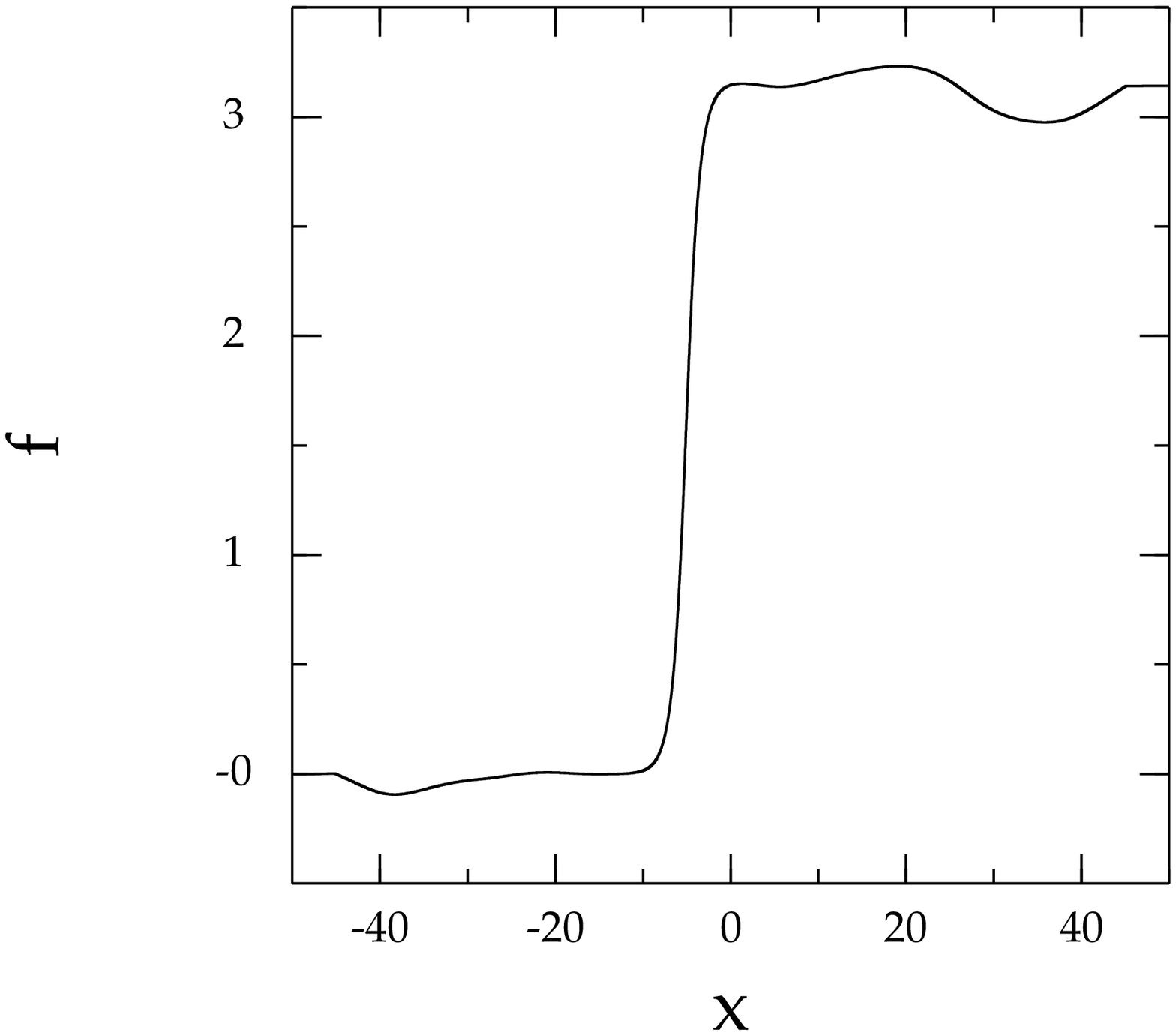}
\end{minipage} 
\caption{Trajectory of a breather sent with velocity $v=0.06587375$
towards a well of width $L=10$ and depth $0.8$ ({\it ie} $\lambda=0.8$); $t=600$ and $t=800$, respectively}
\end{figure}

\section{Further solutions (wobbles etc).}

However, the kinks and breathers are not the only finite energy solutions
of the sine-Gordon model. In fact, the model possesses also solutions
which describe bound states of kinks and breathers. One of such 
solutions, the `wobble' was recently discussed in detail by
K\"alberman \cite{Kalberman}.

 In his paper K\"alberman presents an analytic form of this
solution, shows that it describes a static kink in interaction with
a breather, and then discusses some of its 
properties. 

Recently \cite{laf}, we have looked at more general solutions of the sine-Gordon model
describing kinks and breathers. Our work was based on  the Hirota method \cite{hirota} 
of deriving such solutions. Using this method we obtained 
\begin{equation}
\phi = 2\; {\rm ArcTan}\;\frac{\left[ 2\,\{{\rm
      cotan}\theta\}\,\cos \Gamma_I +   
e^{{\tilde \Gamma}_3}\{ e^{-\Gamma_R} + 
 \rho^2\; e^{\Gamma_R}\}\right]}{\left[ \{e^{-\Gamma_R}
 +   e^{\Gamma_R} \} 
- 2\,\{{\rm cotan}\theta\}\,\rho\, e^{{\tilde \Gamma}_3} \,
\cos\{\Gamma_I+\varphi\}\right]}, 
\label{kink+breather}
\end{equation}
where
\begin{equation}
{\tilde \Gamma_3} =  \gamma_K \{ x-v_K\, t\} + \eta_K
\qquad \qquad \gamma_K =  \cosh \alpha_K
\qquad \qquad v_K =  \tanh \alpha_K, 
\end{equation}
\begin{equation}
\rho =\frac{\cosh\{\alpha_B-\alpha_K\}-\, \cos
  \theta}{\cosh\{\alpha_B-\alpha_K\}+\, \cos \theta} 
\qquad \hbox{and}\qquad
\varphi= 2\, {\rm ArcTan}\, \frac{\varepsilon\,\sin
  \theta}{\sinh\{\alpha_B-\alpha_K\}}. 
\end{equation}
and 
\begin{eqnarray}
\Gamma_R &=& \frac{1}{\sqrt{1-v_B^2}}\, \cos \theta \; \{
x - v_B\, t\}+ \eta_B \nonumber\\  
\Gamma_I &=& \frac{1}{\sqrt{1-v_B^2}}\,\sin \theta \{ 
t -v_B\, x\} +\xi_B.\nonumber
\end{eqnarray}

This corresponds to the most general kink/breather field configuration in which the kink and the breather
have arbitrary velocities (and so move with respect to each other). 
If we now take
$$
\eta_K=\eta_B=v_B=v_K=0 \qquad \quad
\xi_B=\frac{\pi}{2}  
$$
and denote
$$
\omega= \sin \theta \qquad -\frac{\pi}{2}\leq \theta \leq \frac{\pi}{2}
$$
then
\begin{equation}
\rho =\frac{1-\, \sqrt{1-\omega^2}}{1+\,
  \sqrt{1-\omega^2}} 
\qquad \qquad
\varphi= \pm \, \pi. 
\end{equation}
Hence our expression becomes
\begin{equation}
\phi = 2\; {\rm ArcTan}\;\frac{\left[
    \frac{\sqrt{1-\omega^2}}{\omega}\,\sin \{\, \omega\, t\}  +   
\frac{1}{2}\, e^{\, x}\{ e^{-\,\sqrt{1-\omega^2}\, x} + 
 \rho^2\; e^{\,\sqrt{1-\omega^2}\,x}\}\right]}{\left[  \cosh\{
    \, \sqrt{1-\omega^2}\,x\} 
+ \frac{\sqrt{1-\omega^2}}{\omega}\,\rho\, e^{ \, x} \,
\sin\{\omega\, t\}\right]} 
\label{staticbreather}
\end{equation}
where $\omega$ is a frequency varying from $-1$ to $1$.
This agrees with the expression given by K\"alberman which describes a stationary field
configuration in which the kink and the breather sit on `top of each other' and are not in 
relative motion.

As is clear from (\ref{staticbreather}) the field configuration
depends on one parameter (the frequency 
of the breather) and so, as we showed in \cite{laf}, we have studied the stability of this field
configuration. To do this we  
 calculated $\phi$ and its time derivative from (\ref{staticbreather}) and then used them
 as the initial condition for our simulations.
 The results of our
simulations were in complete 
agreement with the analytical expression thus showing that the
solution is stable with respect 
to small perturbations (due to the discretisations).

Next we studied the stability of the wobble with respect to
larger perturbations. 
In particular we changed the original slope of the kink: ({\it i.e.} in the
expression (\ref{staticbreather})  
we replaced $exp(x)$ by $exp(\lambda x)$ where $\lambda\ne1$). As discussed
in \cite{laf} we performed several simulations 
with $\lambda$ ranging from 1.05 to $1.3$.

Our initial conditions corresponded to incorrect
initial field configurations which added some extra energy to the system. The system
was then unstable and so it evolved towards a stable wobble emitting
some radiation which was sent out towards the boundaries of the
grid.  For $\lambda$ close
to one - the perturbations were small - hence the system returned to
its initial configuration 
(with $\lambda=1.0$). For larger values of the perturbation the system
was more perturbed and often  
not only kept on sending out its excess of energy but also,
at regular intervals, altered its frequency of oscillation (increasing
it) which allowed it to 
send out even more radiation.
In fig 4. we present the plots of the time dependence of the total
energy 
 as seen in the simulation in which $\lambda$ was set at 1.15.


\begin{figure}
\begin{center}
\includegraphics[width=14pc]{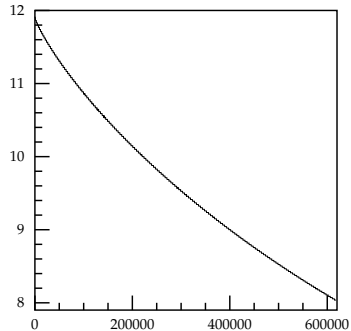}
\end{center}
\caption{\label{label2}Energy as a function of time as seen in a simulation started with 
$\lambda=1.15$}
\end{figure}


Next we performed a series of simulations in which the initial (`wobble') 
configuration  was sent towards a potential hole. As for a single breather \cite{new}
we have found many possibilities. We have also found that the hole can separate the breather
 from the kink (in one simulation 
we even saw the kink being trapped in the hole while the breather bounced off the kink 
 and returned to where the wobble originally came from).
Our results on this are still  preliminary as we have decided to postpone the full study of this
problem to some future work.

We have also looked at field configurations involving a kink with more breathers.
In particular in \cite{laf} we gave an explicit expression for such a configuration
corresponding to a kink and two breathers. We have also verified that this 
configuration is  stable.

\section{The energy}

As we mentioned before the energies of the kink or the breather configurations 
have very simple forms. This has led us to enquire whether this is also the case
for the more general field configurations mentioned above. In fact, we managed to show this: the energies 
of all field configurations that can be derived by the Hirota method are additive 
as they are determined entirely by the asymptotic values of the $\tau_i$ functions
that arise in the construction of these field configurations. 
The $\tau_i$ functions arise
when one sets $\phi= 2\log \frac{\tau_1}{\tau_0}$ and then solves the relevant equations (more details
in \cite{laf}).
Any interested reader can find more detail in \cite{laf}.

In the cases we have discussed here the Hamiltonian density is given by
\begin{equation}
H\,=\,(\partial_t\phi)^2\,+\,(\partial_x\phi)^2\,+\,
\left[\sin\left(\phi\right)\right]^2,       
\label{ham}  
\end{equation}
and so the energy becomes
\begin{equation}
E=\int_{-\infty}^{\infty} dx\, H = 
2\partial_x\left(\ln \tau_0\,+\,\ln
\tau_1\right)\mid_{x=-\infty}^{x=\infty} 
\label{resu}
\end{equation}
and so is determined entirely by the asymptotic values of $\tau_i$ functions. 

In this way we have managed to show that the energies of solutions we have considered in this talk
are:

\begin{enumerate}
\item For the $1$-kink: 
\begin{equation}
E_{1{\rm -soliton}} = 8\frac{1}{\sqrt{1-v^2}}.
\end{equation}
\item For the breather 
\begin{equation}
E_{{\rm breather}} = 16\frac{\sqrt{1-\omega^2}}{\sqrt{1-v^2}}. 
\end{equation}
\item For the wobble
\begin{equation}
E_{{\rm wobble}} =8\frac{1}{\sqrt{1-v_K^2}}+16
\frac{\sqrt{1-\omega^2}}{\sqrt{1-v_B^2}}.  
\end{equation}
\item For a solution involving the kink and two breathers  mentioned above (with their
  velocities set to zero)
\begin{equation}
E_{{\rm kink+2 breathers}} =8+
 16\sqrt{1-\omega_1^2} +
16 \sqrt{1-\omega_2^2}, 
\end{equation}
where $\omega_i=\sin \theta_i$, $i=1,2$. 
\end{enumerate}

\section{Perturbed Field Configurations}
Given that the sine-Gordon model possesses many solutions which resemble perturbed kinks ({\it i.e.} which are given
by kinks and breathers) we have also tried to see what happens when one perturbs a kink and lets it evolve in time.
Again we have looked at various perturbations, paying particular attention to configurations which were generated by
adding to the kink an extra perturbation of the form
\begin{equation}
\label{pert}
\delta\phi(t=0)\,=\,\frac{B}{\cosh(\mu x)},\quad \delta\frac{\partial\phi}{\partial t}(t=0)\,=\,\frac{A}{\cosh(\nu x)}.
\end{equation}
In our simulations we used various values of $A$, $B$, $\mu$ and $\nu$. In all cases the perturbation made the kink
move and generated many moving breather-like configurations. To see what the system would finally settled at
we absorbed the energy at the boundaries of the grid. This had the effect of slowing down the kink and also
of absorbing and/or altering some breather-like structures. The process was very slow and
the results were somewhat inconclusive.
What we can say at this stage is that a general field configuration gradually splits into moving kinks and breathers,
and some radiation, which quickly moves out to the boundaries. However, the resultant field 
configuration is metastable; it still radiates, albeit very slowly, and gradually evolves towards
a field configuration involving mainly a kink. Whether at the end of its evolution we end up with a kink 
or a kink with some breathers is hard to determine.


\section{Simple models}

Here we mention briefly our simple models which partially explain what we have seen
in our numerical simulations.
First we consider the sigma model in 2 dimensions. In this case the soliton possesses
many vibrational modes \cite{baby2} and so we have constructed  a model which treats the
soliton as a system of four masses (connected to each other by springs) \cite{sigma}.
The system of four masses is sent towards the potential hole and then as it falls into it the masses begin to oscillate.
These oscillations then model the soliton vibrations seen in full simulations. The energy is transferred
to these oscillations and if the energy of the centre of mass is too low 
the system is trapped in the hole.
Sometimes, when the system reaches one of the edges of the hole it happens to be in a state that allows the energy of the oscillations
 to get transferred back to the system as a whole (the energy of its centre of mass)
and the soliton can come out. Whether this happens or not depends on the flow 
of the energy between the variational modes and on the kinetic and potential energy of the soliton.
Hence, as we showed in \cite{sigma}, the model reproduces quite well the main features of the scattering pattern 
seen in the full simulations.

In  the Sine-Gordon case we have looked at the old results on the scattering of kinks
on point inpurities \cite{older} (showing a similar trapping/transmission/reflection 
pattern)  and their recent explanation by Goodman and collaborators \cite{Goodman}. 
In \cite{Goodman} the authors explained the observed results by invoking an interaction of 
the kink with the oscillation of the vacuum (around the impurity) which was described
by a standing wave whose amplitude was a further degree of freedom (in addition to the 
position of the kink). 
 The model  reproduced all the features 
of the results of the original simulations reported in \cite{older} 
and so the two models discussed by us in \cite{sine} are based on the adaptation of the
 ideas of Goodman et al to our case. In both models we introduced degrees of freedom describing various standing
waves in the hole (in one model the waves were restricted to the edges of the hole
and in the other they described the global standing waves in the hole).  The waves in both models were chosen
 somewhat arbitrarily as the idea was not to reproduce the pattern in any detail
but just to check whether the mechanism of Goodman can be applied to our case too.

In fact both models worked surprisingly well. They reproduced the pattern quite well,
although the critical value of velocity was a little too high. Given that these models
involved only very few (3 or 4) degrees of freedom we were very encouraged by these results; they
require further work to understand better which modes are important and which are less so.

\section{Final comments and Conclusions.}
We started this talk with a very brief review of the topological solitons and of their dynamics.
Then we reviewed the results of our studies of the scattering of topological solitons 
on a potential obstruction, of both a barrier and a hole-type. We finished by reporting some results 
on breathers and wobbles (bound states of breathers and kinks).

Our results have shown that when a soliton was sent towards the barrier 
its behaviour resembled that
of a point particle. Thus at low energies the soliton was reflected by the barrier
and at higher energy it was transmitted. The scattering process was very
elastic.
During the scattering the kinetic energy of the soliton was gradually
converted into the energy needed to `climb the barrier'. If the soliton
had enough energy to get to the `top' of the barrier then it was transmitted,
otherwise it slid back regaining its kinetic energy.

In the hole case, the situation was different. This time, as the soliton entered
 the hole it gained extra energy. Some of this
energy was converted into the kinetic energy of the soliton, some was
radiated away.  So when the soliton tried to `get out' of the hole it
had less kinetic energy than at its entry and, when this energy was
too low, it remained trapped in the hole.  During the scattering
process, like in the case of a barrier, the soliton's size changed and
so it started oscillating. Afterwards, even when the soliton left the hole, its size continued to oscillate.
 Hence some energy was transferred to the oscillations resulting in the 
emission of radiation  {\it i.e.} in an inelastic
process.

We have also looked at the scattering of breathers on potential holes. As 
breathers can be thought of as bound states of kinks and antikinks 
they can split leaving a trapped kink or an antikink in the hole and allowing
its partner to escape either forwards or backwards. In addition,  as the energy
of the breather depends on the frequency of its oscillation (`breathing') 
this frequency can change as well.  All such phenomena were seen in our
simulations and we hope to report on this more fully in the near future \cite{new}. 

We have also looked at some field configurations, which are solutions of the equations
of motion and which describe bound states of kinks and breathers.
 As the energy of each breather depends on its frequency (and vanishes in the limit
of this frequency going to 1) the extra energy, due to these extra breathers, does not have to be very large.
The solutions appear to be stable and this stability is guaranteed by the integrability of the model.
We have tested this numerically and have found that small perturbations, due to the discretisations, do not
alter this stability. To change it we need something more drastic - like the absorption or the space variation 
of the potential ({\it i.e.} the coefficient of the $sin^2$ term in the Lagrangian). But even then 
the effects are not very large - one sees splitting of breathers {\it etc} but no `global annihilation'.

Finally, all our results (on the scattering of topological solitons on obstructions) 
generalise to other models; such as {\it i.e.} 
the Landau-Lifschitz model with a position dependent potential or an external magnetic 
field, other models in (1+1) dimensions, such as a $\lambda \phi^4$ model
or even models describing ferro- and anti-ferro-magnets \cite{speight}.

Clearly a lot of work still has to be done in this area.

\ack
This talk was given by WJZ at the QTS5 conference which was held in Valladolid in July 2007.
WJZ wants to thank the Organisers for their hospitality.


{}

\end{document}